\documentclass[%
 aip,
 amsmath,amssymb,
 reprint,%
]{revtex4-1}

\usepackage{graphicx}
\usepackage{dcolumn}
\usepackage{bm}

\usepackage[utf8]{inputenc}
\usepackage[T1]{fontenc}
\usepackage{mathptmx}

\begin{document}

\preprint{AIP/123-QED}

\title{A hybrid ion-atom trap with integrated high resolution mass spectrometer}

\author{S Jyothi}
\affiliation{Department of Electrical and Computer Engineering, Duke University, Durham, North Carolina, USA}
\affiliation{School of Chemistry and Biochemistry, Georgia Institute of Technology, Atlanta, Georgia, USA}

\author{Kisra  N Egodapitiya}
\affiliation{Department of Electrical and Computer Engineering, Duke University, Durham, North Carolina, USA}
\affiliation{School of Chemistry and Biochemistry, Georgia Institute of Technology, Atlanta, Georgia, USA}
\author{Brad Bondurant} 
\affiliation{Department of Electrical and Computer Engineering, Duke University, Durham, North Carolina, USA}
\author{Zhubing Jia}%
\affiliation{Department of Physics, Duke University, Durham, North Carolina, USA}
\affiliation{School of Physics, Georgia Institute of Technology, Atlanta, Georgia, USA}
\author{Eric Pretzsch} 
\affiliation{School of Physics, Georgia Institute of Technology, Atlanta, Georgia, USA}
\author{Piero Chiappina} 
\affiliation{School of Physics, Georgia Institute of Technology, Atlanta, Georgia, USA}
\author{Gang Shu} 
\affiliation{School of Chemistry and Biochemistry, Georgia Institute of Technology, Atlanta, Georgia, USA}
\author{Kenneth R Brown} 
\email{ken.brown@duke.edu}
\affiliation{Department of Electrical and Computer Engineering, Duke University, Durham, North Carolina, USA}
\affiliation{School of Chemistry and Biochemistry, Georgia Institute of Technology, Atlanta, Georgia, USA}
\affiliation{Department of Physics, Duke University, Durham, North Carolina, USA}
\affiliation{School of Physics, Georgia Institute of Technology, Atlanta, Georgia, USA}
\affiliation{Department of Chemistry, Duke University, Durham, North Carolina, USA}
\affiliation{School of Computational Science and Engineering, Georgia Institute of Technology, Atlanta, Georgia, USA}
%
\date{\today}

\begin{abstract}
In this article we describe the design, construction and implementation of our ion-atom hybrid system incorporating a high resolution time of flight mass spectrometer (TOFMS). Potassium atoms ($^{39}$K) in a Magneto Optical Trap (MOT) and laser cooled calcium ions ($^{40}$Ca$^+$) in a linear Paul trap are spatially overlapped and the combined trap is integrated with a TOFMS for radial extraction and detection of reaction products. We also present some experimental results showing interactions between $^{39}$K$^+$ and $^{39}$K, $^{40}$Ca$^+$ and $^{39}$K$^+$ as well as $^{40}$Ca$^+$ and $^{39}$K pairs. Finally, we discuss prospects for cooling CaH$^+$ molecular ions in the hybrid ion-atom system. 
\end{abstract}

\maketitle


\section{\label{sec:level2} Introduction}

In the past decade, hybrid traps were employed to study elastic and inelastic collisions as well as rich chemical interactions between cold atoms and ions \cite{Grier2009,Zipkes2010,Schmid2010,Hall2011,Rellergert2011,Harter2012,Ravi2012,Sivarajah2012,Ray2014,Haze2015,Meir2016,Hall2012,Deiglmayr2012,Rellergert2013,Kwolek2019}. The very recent research in this field focuses on controlling the chemical reactions in cold ion-atom collisions. These include controlling the electronic state \cite{Joger2017,Saito2017}, the spin state \cite{Sikorsky2018} or the collision energy \cite{Saito2017,Puri2018}. The long interrogation time and precise internal state control make these systems an ideal platform to explore quantum state dependent interactions. 

Ion-atom hybrid traps are also proposed to attain the rovibrational ground state of molecular ions \cite{Hudson2016}. The rich internal structure of molecules makes them well suited for precision experiments such as time variation of fundamental constants \cite{schiller2005}, precise measurement of the electron’s electric dipole moment \cite{William2017} and study of parity violation \cite{Kozlov1995}. Laser cooled atomic ions have proved to efficiently cool the translational degrees of freedom of co-trapped molecular ions \cite{kimura2011,Rugango2015}. The idea is to cool the internal states of the molecules sympathetically using ultra cold atoms. Recent experiments to achieve this goal gave promising results. Vibrational cooling of BaCl$^+$ ions \cite{Rellergert2013} was demonstrated using interaction with neutral Ca atoms in a MOT. Rotational cooling of MgH$^+$ down to 7.5 K was achieved with helium buffer gas cooling \cite{Hansen2014}. One can expect more efficient rotational cooling using interactions with laser cooled atoms. This technique is independent of the molecular ion species as it does not involve any resonance phenomena \cite{Hudson2016}. One of the goals of our ion-atom hybrid trap presented in this manuscript involves realizing CaH$^+$ molecular ions in their internal and external ground state. 

In collision experiments involving different species, a reliable detection scheme to identify the reaction products is essential. In the past, indirect detection either by secular excitation of the trapped ions \cite{Baba2002} or structural changes in the ion coulomb crystal, in conjunction with molecular dynamics simulations \cite{Roth2006}, were used to identify the reaction products. Even though these indirect methods simplify the equipmental design, they lack high resolution and require more sophisticated data analysis. A time of flight mass spectrometer (TOFMS) can be coupled to an ion trap for axial \cite{Ravi2012APB,SECK2014} or radial \cite{Schowalter2012,Meyer2015,jyothi2015,Deb2015,Rosch2016,Heather2017} extraction and detection of the trapped ions. The radial extraction has proven to be advantageous in improving the mass resolution because the trapped ion spatial extent is smaller along the radial direction. In our ion-atom hybrid trap, we have incorporated a TOFMS to radially extract the trapped ions with high mass resolution. 

In this article we describe the design and implementation of our ion-atom hybrid trap which incorporates a MOT for potassium atoms ($^{39}$K) and a linear Paul trap for ions. In section II, the simultaneous cooling and trapping of $^{39}$K and $^{40}$Ca$^+$ as well as the details of TOFMS are discussed. The details of experimental control is also described in this section. In section III, we present measurements to demonstrate the spatial overlap of ion-atom trap centers and a few experimental results to demonstrate the potential of the experimental system to perform cold interactions studies. These include elastic as well as resonant charge exchange collisions between $^{39}$K$^+$ and $^{39}$K, long range Coulomb interaction between $^{40}$Ca$^+$ and $^{39}$K$^+$, and charge transfer collisions between $^{40}$Ca$^+$ and $^{39}$K. We conclude by discussing future goals for the hybrid system.  

\section{\label{sec:level2} Experimental Design}

\begin{figure}[t]
\includegraphics[scale=0.7]{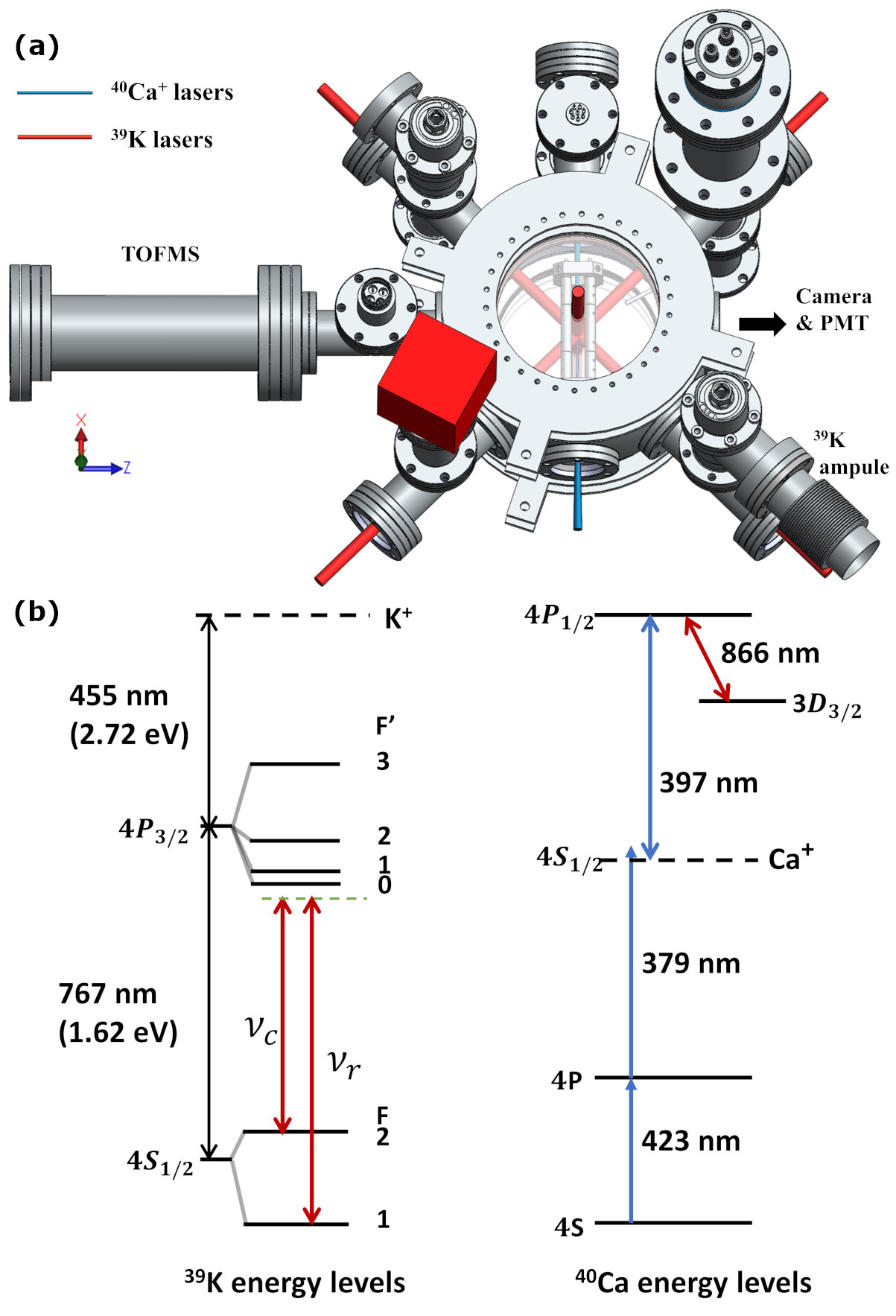}
\caption{\label{fig:expt} Schematic of the experimental set up. (a) Top view of the experimental set up illustrating the atom trap, ion trap and the TOFMS. (b) The energy levels of $^{39}$K and $^{40}$Ca$^+$ (not to scale) relevant for our experiments.}
\end{figure}

The experimental setup consists of an 8$''$ octagon vacuum chamber (Kimball Physics MCF800-SphOct-G2C8) maintained at a pressure of 1$\times$10$^{-9}$~Torr. A schematic of the experimental system is illustrated in Fig. \ref{fig:expt} (a). A segmented electrode linear quadrupole trap (LQT) is mounted at the center of the chamber. The ion trap consists of four rods in a quadrupole configuration. Each rod is broken into three electrodes. A radio frequency (rf) electric field is applied on the two central diagonal electrodes to confine the ions in the radial directions (YZ). The other two central electrodes function as the rf ground. The eight end-cap electrodes are used to apply direct current (dc) voltages to confine the ions in the trap axial direction (X). The rods are assembled such that the neighboring rod separation is 10~mm in order to accommodate the 10~mm MOT beams. The experimental chamber is integrated with a high-resolution TOFMS for the identification of the ionic species. The TOFMS is mounted to one of the 2.75$''$ ports perpendicular to the ion trap axis for radial ejection of the ions. The ion trap voltages and fast rf switching for the TOFMS are performed using the electronics developed by Hudson Lab at UCLA \cite{Schneider2014}. The chamber is also equipped with a leak valve to input gases such as H$_2$ for creating molecular ions for future experiments.

The $^{39}$K MOT is formed using three mutually orthogonal retro-reflected laser beams. The ground state hyperfine splitting of 4$^2$S$_{1/2}$ of $^{39}$K isotope is 461.7~MHz and the excited state manifold is just 33.5~MHz. This hyperfine structure makes it possible to derive both the cooling and the repumper laser beams using a single laser at 766.7 nm (external cavity diode laser and tapered amplifier, MOGLabs). The laser is locked to the crossover peak of the saturation absorption spectrum. The laser output is split into two beams and are sent through acousto-optic modulators (AOM) to get the desired detunings of 20 MHz from the F=2 (cooling) and F=1 (repumper) to the excited state manifold. Potassium vapor is produced by heating a potassium ampule (Sigma Aldrich 244856) placed in a stainless steel bellow. The glass ampule is connected to the chamber via an all metal right angle valve that can be used to isolate the ampule if required, and is broken during the initial pump down of the chamber. The magnetic field gradient required for the MOT is produced using a pair of coils in anti-Helmholtz configuration, with a field gradient of about 10~G$/$cm in the axial direction. The MOT anti-Helmholtz magnetic coils are symmetrically mounted outside the chamber on a two dimentional translation stage to precisely adjust the center of the magnetic field. An additional pair of coils in Helmholtz configuration is mounted concentric with the MOT coils. These coils are used to shift the magnetic field center in the vertical direction to shift the MOT atoms from the ion trap center when required. 

In our hybrid trap experiments we use laser cooled $^{40}$Ca$^+$ ions. Neutral calcium atoms are produced by evaporating a piece of calcium inside a stainless steel tube by passing a current through a steel wire inserted into the stainless steel tube. The calcium ions are produced by resonant two photon ionization using 423 nm and 379 nm laser beams \cite {Lucas2004}. The ions are trapped with an rf field of 400 V at 1.7 MHz and a dc voltage of 7 V and are cooled by 397 nm laser with 866 nm laser as the repumper. All wavelengths are derived from a multi diode laser module (Toptica Photonics).  

For the simultaneous cooling and trapping of $^{39}$K and $^{40}$Ca$^+$, the 767 nm laser beams for $^{39}$K and the 397 nm cooling beam for $^{40}$Ca$^+$ need to be alternately switched at high frequency. This is because the ionization energy of the potassium atoms from the $^4$P$_{3/2}$ excited state is only 2.7 eV and any light with wavelength lower than 455 nm can ionize the excited state $^{39}$K atoms (the relevant energy levels for $^{39}$K and $^{40}$Ca$^+$ are shown in Fig. \ref{fig:expt} (b)). Hence, while the MOT is operational, the 767 nm and 397 nm lasers are alternately switched at 2 kHz and the 423 nm and 379 nm lasers are blocked using a mechanical shutter to avoid ionization of excited state $^{39}$K atoms. Hence in all experiments, we first load the $^{40}$Ca$^+$ ions, block the 423 nm and 379 nm laser beams, start alternate switching of the 767 nm and 397 nm laser beams using AOMs and start loading the $^{39}$K MOT. 

\begin{figure}
\includegraphics[scale=0.55]{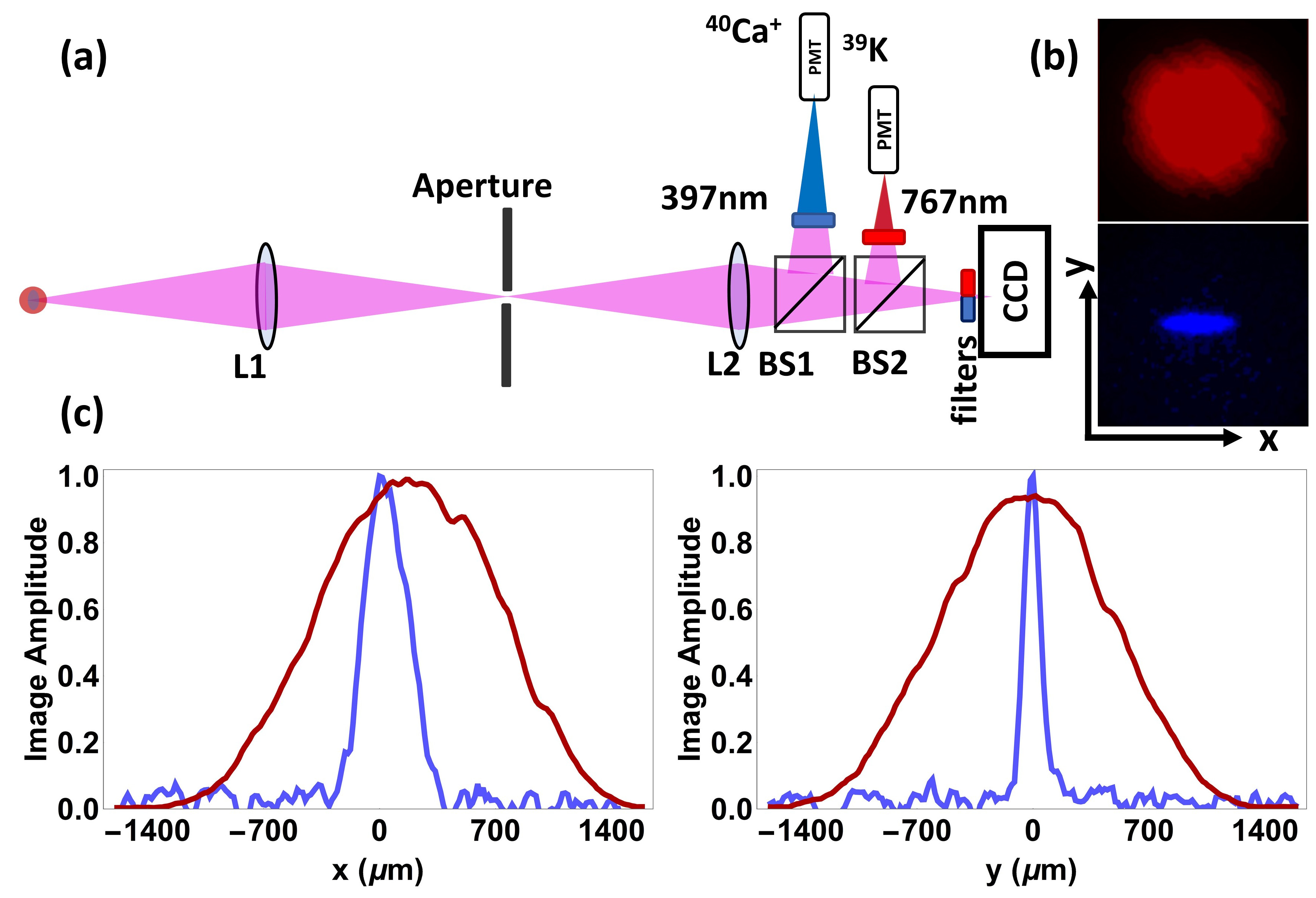}
\caption{\label{fig:overlap_1} Fluorescence detection and ion-atom spatial overlap. (a) Schematic of the spatial filtering set up to collect the $^{39}$K atom and $^{40}$Ca$^+$ ion fluorescence. (b) False color images of $^{39}$K MOT and $^{40}$Ca$^+$ laser cooled ions. (c) The spatial profile of the atoms and ions are plotted in two directions of the image plane (XY) of the camera. Ion-atom spatial overlap in the third dimension is demonstrated in Fig.\ref{fig:overlap_2}.}
\end{figure}

The fluorescence from the $^{39}$K atoms and $^{40}$Ca$^+$ ions are collected using two calibrated photo multiplier tubes (PMT) and an EMCCD camera. A schematic of the fluorescence collection is shown in Fig. \ref{fig:overlap_1}. The camera images are used to confirm the ion-atom spatial overlap in one plane (XY) along the ion trap axis. The spatial overlap in the third direction (Z) cannot be imaged due to design constrains and is ensured by another method described in the result section. The overlap between the atoms and the ions can be optimized by tuning the trapped ion position using the dc voltages on the ion trap electrodes or by tuning the MOT position by shifting the magnetic field center in the horizontal plane using a 2D translation stage and in the vertical axis by adding a constant magnetic field using a Helmholtz coil pair.

\begin{figure}
\includegraphics[scale=0.35]{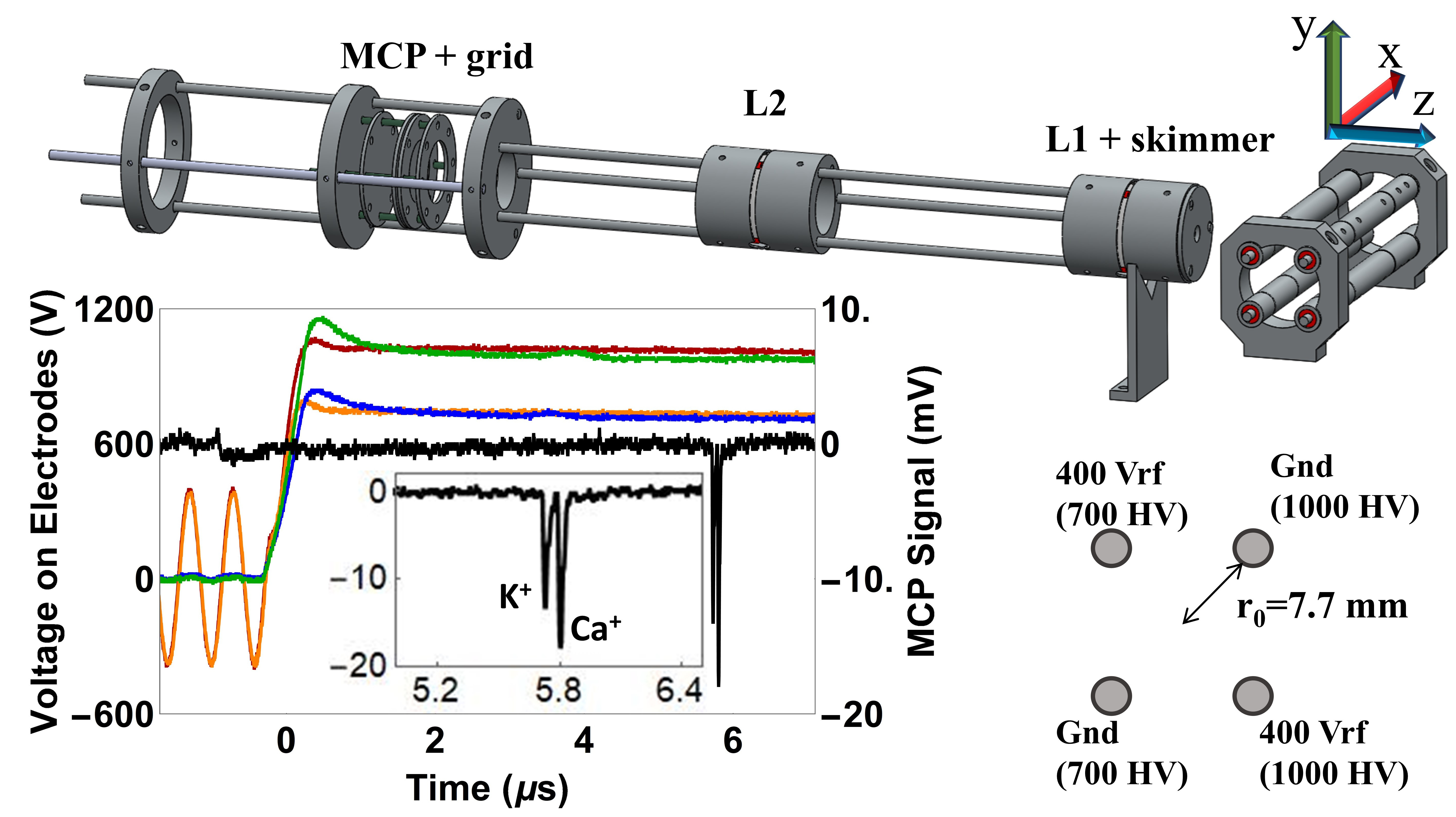}
\caption{\label{fig:TOF} Schematic of the LQT-TOFMS. The ions are radially extracted from the LQT to the TOFMS by switching the voltages on the central four electrodes from rf to dc. The voltages applied on the central rods while trapping is shown and the extraction dc high voltages are shown in brackets. The oscilloscope traces of voltage switching and a representative MCP signal are shown in the graph. The inset shows a zoomed in portion of the TOF mass spectrum with $^{39}$K$^+$ and $^{40}$Ca$^+$ ions.}
\end{figure}

The TOFMS is an essential part of the experimental set up to identify the reaction products of the cold collision experiments, especially the nonfluorescent ions. A schematic of the TOFMS is shown in the Fig. \ref{fig:TOF}. It consists of a grounded skimmer which determines the entrance aperture of the TOFMS (5.6 mm in diameter), two Einzel lenses for ion collimation, a micro channel plate (MCP) for ion detection and a grounded wire mesh in front of the MCP detector to shield the MCP from the drift tube. The entire TOFMS parts are mounted on a 267 mm long drift tube and is attached to the 2.75$''$ ports perpendicular to the ion trap axis for the radial ejection of the ions. The MCP (Photonis) consists of a pair of impedance matched microchannel plates with an active area of 18 mm. The two plates are assembled in a chevron configuration allowing maximum signal multiplication. A metal anode at the back of the MCP detector collects the electrons resulting from charged particle detection. The timing signal is extracted from the back plate of the MCP pair through an RC decoupling circuit and is collected on a digital storage oscilloscope.

The MCP is maintained at a high voltage of -2.1 kV. To extract the ions onto the TOFMS the trapping rf voltages are turned off and appropriate dc high voltages (approximately 1000 V and 700 V) are applied to the 4 central rods to create a potential gradient towards the MCP. To improve the resolution of the time of flight spectra of the ions, the residual oscillation of the rf field after it has been turned off is damped. To facilitate this, the electronics system is equipped with damping circuitry to remove residual rf oscillations. After the ions are ejected, the ions travel through a field free region. In order to collimate the ions in the two planes perpendicular to the spectrometer axis, high voltages are applied to the two Einzel lenses. The construction of the Einzel lens is similar to that used in \cite{Schneider2014}. The lens voltages are adjusted to optimize the TOFMS spectrum. In the present experiment a resolution, m/$\Delta$m of about 208 was achieved. This resolution is sufficient to resolve the $^{39}$K$^+$ and $^{40}$Ca$^+$ ions in our experiments, whose masses differ by only one atomic unit.

\begin{figure}
\includegraphics[scale=0.5]{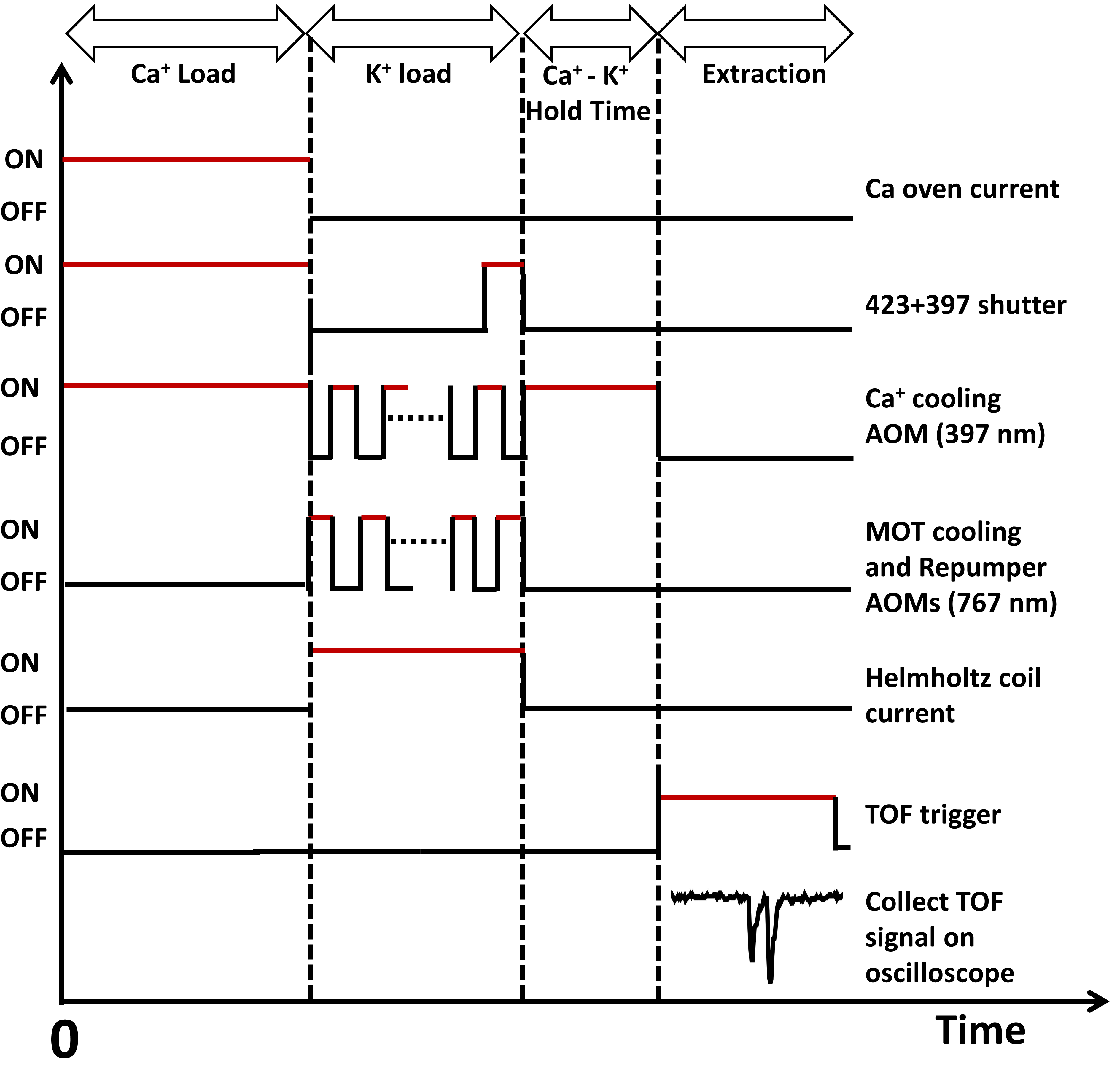}
\caption{\label{fig:timing} Experimental control sequence. An example of the experimental control sequence used to observe the sympathetic cooling of $^{39}$K$^+$ ions by laser cooled $^{40}$Ca$^+$ ions as described in the section III. B.}
\end{figure}

The real-time control of the experiment was implemented using hardware and software solutions provided by the Advanced Real-Time Infrastructure for Quantum physics (ARTIQ) framework \cite{sebastien_bourdeauducq_2018_1492176} from m-labs. Using ARTIQ, experiment code is written in Python, then compiled and executed on dedicated Field Programmable Gate Array (FPGA) hardware with nanosecond timing resolution and sub-microsecond latency. The central controller used in the experiment was the Kasli FPGA carrier, which was accompanied by several Eurocard Extension Modules (EEMs) to complete the setup. Two digital input/output EEMs, each with four channels, were used to count PMT inputs as well as to send trigger signals to other electronics such as the mechanical shutter and TOFMS trigger. A third EEM contained four DDS output channels from an AD9910 DDS chip and was used to drive the AOMs. An example timing sequence for the measurement of sympathetic cooling of $^{39}$K$^+$ ions by laser cooled $^{40}$Ca$^+$ ions as discussed in section III (B) is shown in Fig. \ref{fig:timing}.

\section{\label{sec:level2} Results}

\begin{figure}
\includegraphics[scale=0.55]{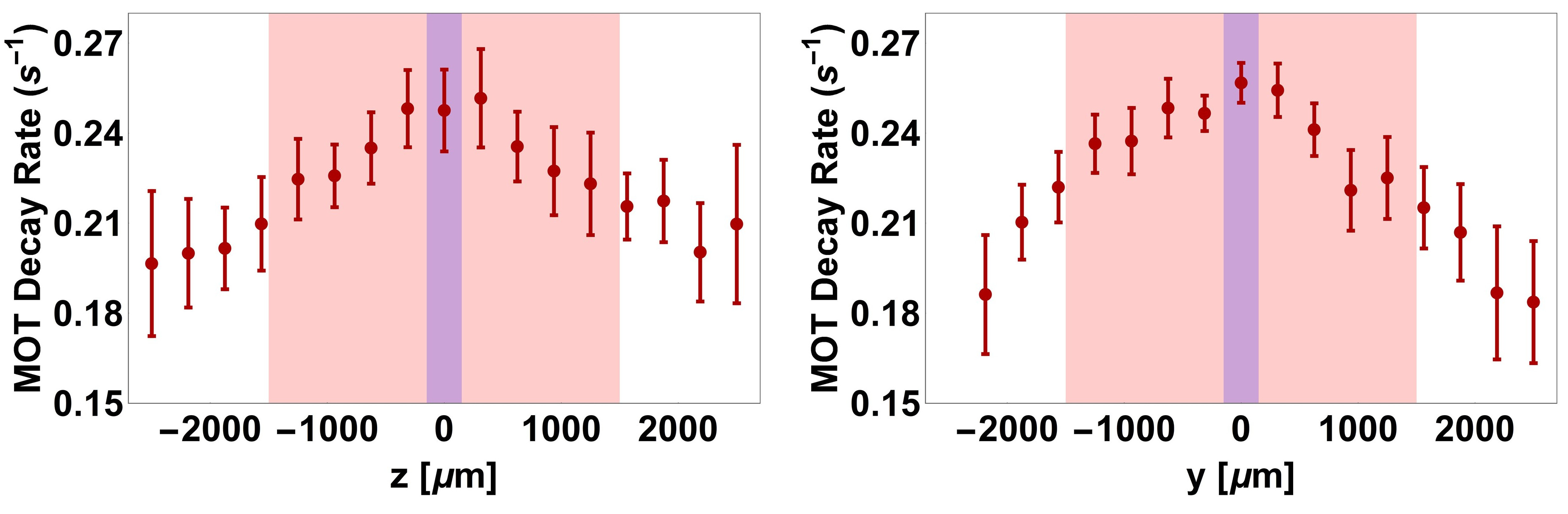}
\caption{\label{fig:overlap_2} Ion-atom spatial overlap. Rate of ionization of $^{39}$K MOT atoms by the cooling laser for $^{40}$Ca$^+$ ions as a function of position of the 397 nm laser along the Y and Z axes (see text for details). The Y=Z=0 represents the ion trap center. The focused axial beam position is scanned in space using a translational stage to ionize the MOT atoms at different positions. The position of the beam to get maximum ion fluorescence is shown by the blue shaded area and the red shaded area represents the MOT spatial extent. The measurement ensures the ion-atom spatial overlap along the Z axis which is not observable using the camera images.}
\end{figure}

The main requirement to effectively use the potential of an ion-atom hybrid trap is the spatial overlap between the trap centers to initiate the interactions. Overlap can be visually observed using the CCD camera if the species are fluorescing. In our set up we have the camera set up along only one axis and hence we can see overlap only in one plane. We measure the overlap in the third direction by using the fact that the $^{39}$K atoms in excited state get ionized by the cooling laser for the $^{40}$Ca$^+$ ions. We use this effect to ensure the ion-atom spatial overlap in the following way. The MOT image analysis shows that the MOT full width is less than approximately 3~mm. The $^{40}$Ca$^+$ cooling laser beam radius at its focus is approximately 150 $\mu$m and the beam position can be scanned spatially across the MOT in the YZ plane using a translation stage. First, the $^{40}$Ca$^+$ axial cooling beam is optimally aligned with respect to the ion trap center by observing the fluorescence of $^{40}$Ca$^+$ ions (denoted as position Y=Z=0). The ions are removed and after loading the MOT to saturation, the 397 nm laser is turned on and the MOT decay rate is measured at Y=Z=0. The 397 nm laser beam is scanned across Y(at Z=0) and Z(at Y=0) directions and the MOT decay rate is measured. The rates as functions of Y and Z as given in Fig. \ref{fig:overlap_2} show that the center of the MOT gaussian profile matches with the ion trap center. The measurement reveals good overlap between the trap centers along the Z axis, which is not observable using the camera. These measurements are done with the ion trap voltages off. For all other experiments, the 767 nm and 397 nm lasers are switched alternatively to avoid the MOT atom ionization. 

In the following three subsections we describe the experimental results of  collision experiments performed in the hybrid trap.

\subsection{Sympathetic cooling of $^{39}$K$^+$ by $^{39}$K MOT}
Like every alkali metal ion, $^{39}$K$^+$ ions have closed electronic structure and cannot be laser cooled due to unavailability of suitable lasers. The established way of cooling such ions is via sympathetic cooling by laser cooled atoms or laser cooled ions. The elastic collisions are more probable in an ion-atom hybrid system, leading to transfer of energy from the hot ions to the cold atoms. However, if the mass of the atoms and ions are the same, resonant charge exchange interaction plays an important role in accelerating the cooling process \cite{Ravi2012,Sivarajah2012,Ray2014}. In a charge exchange collision between $^{39}$K and $^{39}$K$^+$, the electron hopes from $^{39}$K atoms to $^{39}$K$^+$ ions in a glancing collision, giving rise to cold $^{39}$K$^+$ ions and hot $^{39}$K atoms. The hot $^{39}$K atoms are lost from the MOT and the MOT is immediately refilled by another $^{39}$K atom as the MOT is loaded continuously.

\begin{figure}
\includegraphics[scale=0.9]{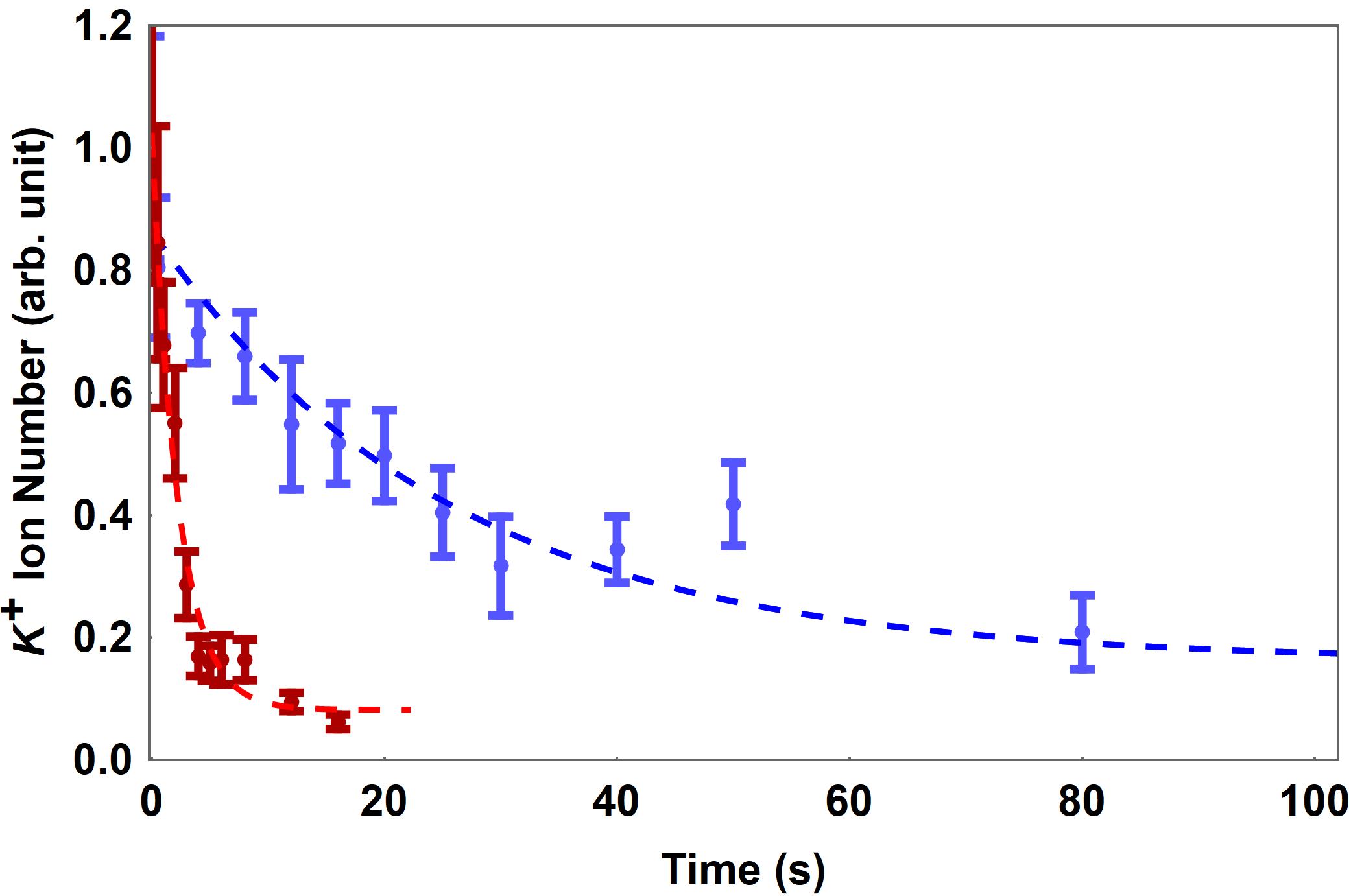}
\caption{\label{fig:K+KMOT} Sympathetic cooling of $^{39}$K$^+$ ions by $^{39}$K MOT. The red squares and blue dots represent the number of $^{39}$K$^+$ ions as a function of hold time in the ion trap in the absence and presence of $^{39}$K MOT atoms, respectively. The dashed lines represent an exponential fit to the data. The $^{39}$K$^+$ ion lifetime is improved by an order of magnitude from 2.2(4) s to 25.3(4) s in the presence of $^{39}$K MOT atoms as a result of elastic as well as resonant charge exchange collisions.}
\end{figure}

Fig. \ref{fig:K+KMOT} demonstrates the cooling of $^{39}$K$^+$ ions by $^{39}$K MOT atoms. In this experiment, after loading the MOT to saturation, $^{39}$K$^+$ ions are produced by photoionizing $^{39}$K atoms in the MOT for 10 ms by the 397 nm laser. The $^{39}$K$^+$ ions are held in the ion trap for fixed amount of hold times either in the absence or in the presence of $^{39}$K atoms in the MOT and then extracted onto the TOFMS. The life time of the $^{39}$K$^+$ ions in the absence of cooling is only 2.2(4) s due to heating caused by trap imperfections, rf heating as well as background collisions. However when held with the $^{39}$K MOT, the lifetime of $^{39}$K$^+$ ions increases to 25.3(4) s. The cooling caused by resonant charge exchange as well as elastic collisions increased the ion lifetime approximately by an order of magnitude.

\subsection{Sympathetic cooling of $^{39}$K$^+$ by $^{40}$Ca$^+$}

The long range Coulomb interaction between co-trapped ions leads to the sympathetic cooling of hot atomic or molecular ions by laser cooled ions \cite{Larson1986}. Here we demonstrate the sympathetic cooling of $^{39}$K$^+$ ions by laser cooled $^{40}$Ca$^+$ ions.

\begin{figure}
\includegraphics[scale=0.9]{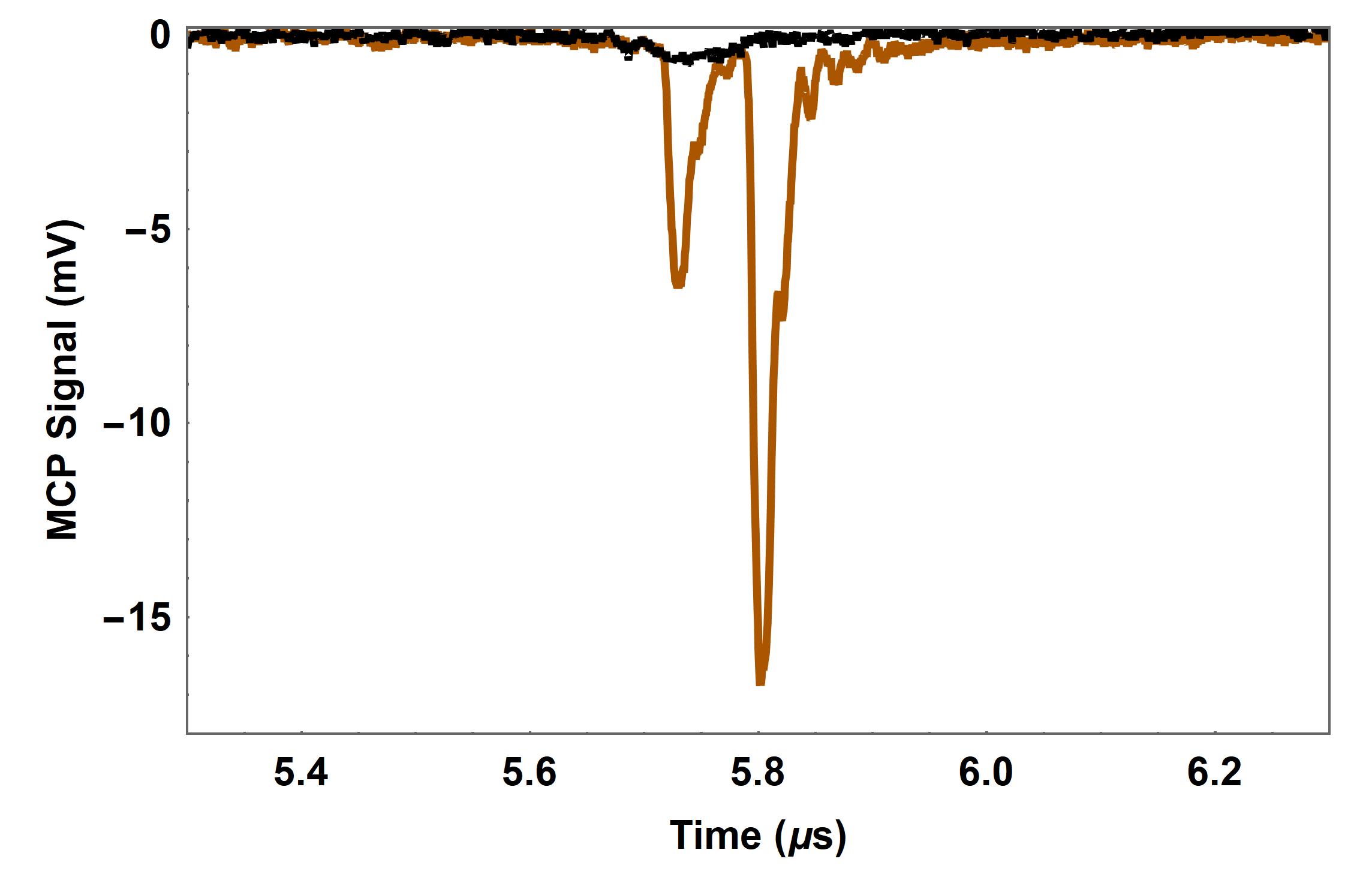}
\caption{\label{fig:TOF_KPlus_CaPlus} Sympathetic cooling of $^{39}$K$^+$ ions by laser cooled $^{40}$Ca$^+$ ions. The black curve represents the potassium TOFMS peak in the absence of $^{40}$Ca$^+$ ions. When co-trapped with the laser cooled $^{40}$Ca$^+$ ions, the sympathetic cooling leads to improved mass resolution (m/$\Delta$m = 208) as shown by the orange curve. The $^{39}$K$^+$ peak height is increased from 0.7 mV to 6.5 mV in the presence of the cold $^{40}$Ca$^+$ ions.}
\end{figure}

The experimental sequence is shown in Fig. \ref{fig:timing}. First, the $^{40}$Ca$^+$ ions are loaded to the ion trap. The MOT is then loaded a few millimeters above the ion trap center and are photoionized to create $^{39}$K$^+$ ions. The trap centers are shifted to avoid any ion-atom charge exchange interaction between $^{39}$K atoms and $^{40}$Ca$^+$ ions. $^{40}$Ca$^+$ and $^{39}$K$^+$ masses are very close and can be trapped with the same optimal trapping parameters. The ions are then extracted onto the TOFMS. Fig.\ref{fig:TOF_KPlus_CaPlus} shows an average of 10 TOFMS traces of $^{39}$K$^+$ in the absence and presence of laser cooled $^{40}$Ca$^+$ ions. The effect of the sympathetic cooling can be observed in the time of flight spectrum as an increase in height and a decrease in width of the $^{39}$K$^+$ peak in the presence of laser cooled $^{40}$Ca$^+$ ions \cite{Schneider2014}. The mass resolution increases from 26 to 208 in the presence $^{40}$Ca$^+$ ions. The $^{39}$K$^+$ peak height also is increased by approximately an order of magnitude from 0.7 mV to 6.5 mV in the presence of the laser cooled $^{40}$Ca$^+$ ions. 

\subsection{Charge exchange collision between $^{40}$Ca$^+$ and $^{39}$K}

\begin{figure}
\includegraphics[scale=0.9]{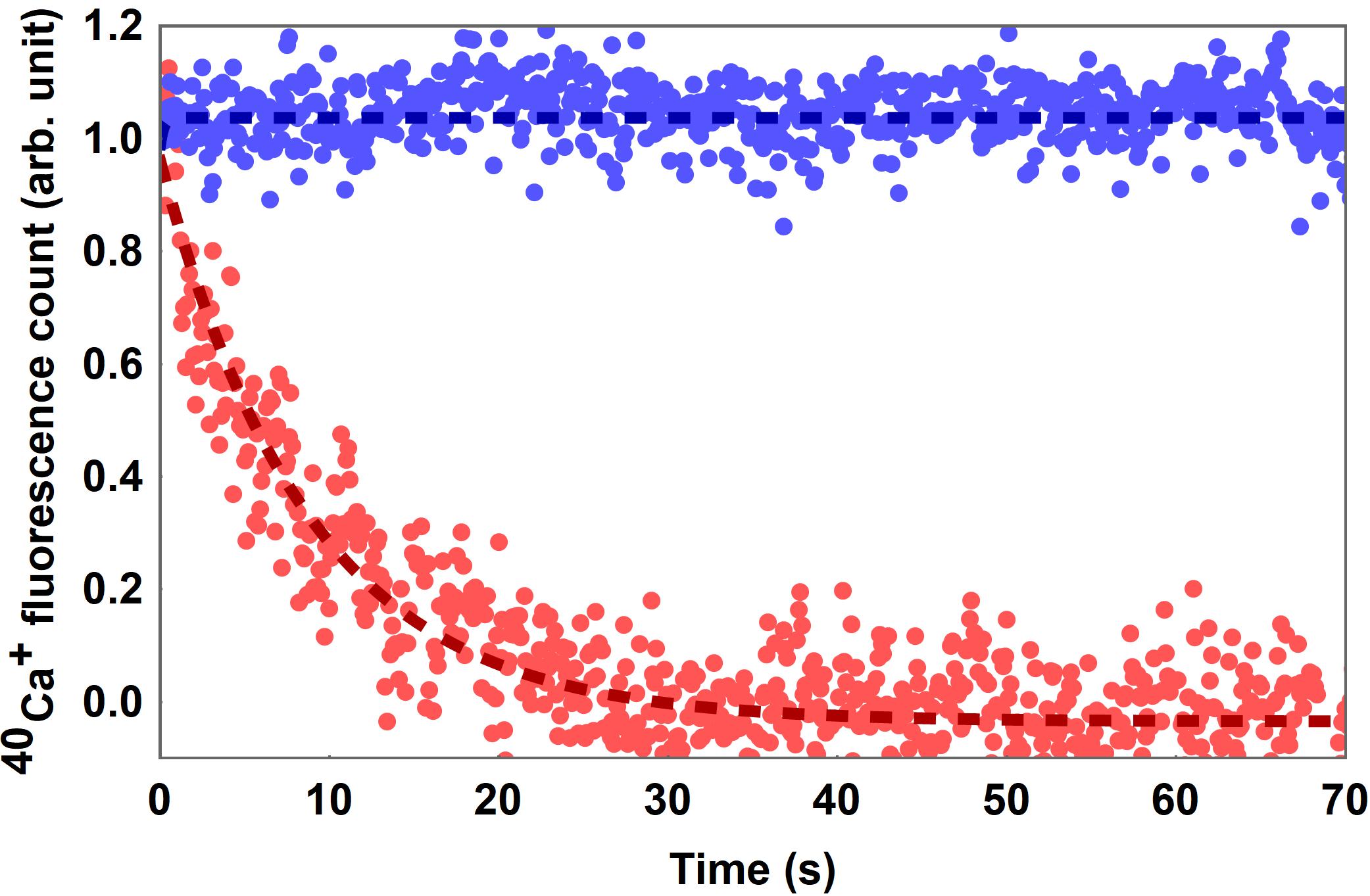}
\caption{\label{fig:KandCa+} Charge exchange collisions between $^{39}$K MOT atoms and $^{40}$Ca$^+$ ions. The fluorescence collected from the $^{40}$Ca$^+$ ions as a function of time is plotted. The blue and red dots represent the lifetime of $^{40}$Ca$^+$ ions in the absence and presence of $^{39}$K MOT atoms, respectively. The dashed lines represent an exponential fit to the data. The laser cooled $^{40}$Ca$^+$ ions lifetime is a few minutes in the absence of MOT. But when held with the $^{39}$K MOT atoms, the lifetime reduces to 8.7(5) s.}
\end{figure}

The long range interaction between neutral atoms and ions \cite{cote2000}scales as $1/r^4$. The details of the long range potential and hence the charge exchange probabilities varies with the combination of atoms and ions \cite{Joger2017,Sikorsky2018,Kwolek2019,Saito2017}. We observe for the first time the charge exchange interaction between $^{39}$K atoms and $^{40}$Ca$^+$ ions. The charge exchange rate can be measured by collecting the $^{40}$Ca$^+$ fluorescence in the absence and presence of $^{39}$K atoms as a function of time and is shown in Fig.\ref{fig:KandCa+}. Alternately, we can observe the reaction as a decrease in the $^{40}$Ca$^+$ ion count and a simultaneous increase in $^{39}$K$^+$ ion count using the time of flight measurements. The lifetime of laser cooled $^{40}$Ca$^+$ ions is a few minutes in the absence of $^{39}$K atoms, but when spatially overlapped with the $^{39}$K MOT, the $^{40}$Ca$^+$ ion number decreases due to charge exchange interaction. The $^{40}$Ca$^+$ ions decay in 8.7(5) s with a MOT of density of 6.7(6)$\times10^8$  atoms/cm$^3$. The details of this interaction will be presented in a subsequent publication \cite{JyothiKCa}.

\section{\label{sec:level2} Summary and Discussion}
We have presented the design details of our ion-atom hybrid trap integrated with a high resolution mass spectrometer, the experimental control details and the detection schemes. We have also presented the ion-ion and ion-atom interactions measured in the hybrid trap. The next step ahead involves control of these interaction using state preparation techniques.  

In addition, it would be possible to induce strong interaction between the first and the second electronic potentials of KCa$^+$ using an additional high intensity laser. This interaction can lead to non-adiabatic transition between these potentials and even make it possible to observe a light-induced conical intersection in the charge exchange interaction \cite{SvetlanaPC}. 

Attaining the ultimate ground state of molecular ions using interactions with laser cooled atoms and ions can be attempted in ion-atom hybrid traps. This would be an alternative approach to coherent internal state preparation by projection into a pure rotational state \cite{Chou2017}. We are planning to produce CaH$^+$ molecular ions by reacting $^{40}$Ca$^+$ ions in 4P$_{1/2}$ state by H$_2$ molecules leaked into the chamber using a leak valve. Translational cooling of CaH$^+$ by laser cooled $^{40}$Ca$^+$ \cite{Rugango2015} and resolved rovibrational spectroscopy of CaH$^+$ \cite{Calvin2018,Calvin2018b} have already been established in our lab. Our goal is to get the rovibrational ground state of CaH$^+$ using cold $^{39}$K atoms. The resolved rovibrational spectrum of CaH$^+$ can be used for the detection of the internal cooling of CaH$^+$. CaH$^+$ prepared in the ground state is a candidate for measuring the time variation of the proton to electron mass ratio \cite{schiller2005}. 

The K-Ca$^+$ hybrid system described here can be used to create a wide range of cold molecular ions beyond CaH$^+$. Molecular ions with mass-ratios within a factor of 5 of Ca$^+$ can be efficiently sympathetically cooled translationally \cite{Schmidt2012}.  The internal motion of molecular ions can be generically cooled by the $^{39}$K MOT as long as the reactions between $^{39}$K and molecules are relatively slow. The hybrid system is ideal for exploring the physics and chemistry of inelastic collisions between $^{39}$K atoms and polyatomic molecular ions.

\begin{acknowledgments}
The authors acknowledge Eric Hudson's group at UCLA for the technical support on the ion trap electronics. This work was supported by the MURI Army Research Office Grant W911NF-14-1-0378-P00008 and ARO Grant W911NF-17-1-0071.
\end{acknowledgments}
\providecommand{\noopsort}[1]{}\providecommand{\singleletter}[1]{#1}%
%

\end{document}